\documentclass[a4paper]{article}
\usepackage{times}
\usepackage{braket}
\usepackage{amsmath}
\usepackage{amssymb}
\usepackage{rotating}

\begin{document}

\title{A term-rewriting system for computer quantum algebra}
\author{J.J. Hudson\\
The Blackett Laboratory, Imperial College London,\\
Prince Consort Road, London SW7 2AZ. United Kingdom.}
\date{\today}

\maketitle

\begin{abstract}
Existing computer algebra packages do not fully support quantum mechanics calculations in Dirac's notation. I present the foundation for building such support: a mathematical system for the symbolic manipulation of expressions used in the invariant formalism of quantum mechanics. I first describe the essential mathematical features of the Hilbert-space invariant formalism. This is followed by a formal characterisation of all possible algebraic expressions in this formalism. This characterisation is provided in the form of a set of terms. Rewrite rules over this set of terms are then developed that correspond to allowed manipulations of the algebraic expressions. This approach is contrasted with current attempts to build invariant quantum mechanics calculations into computer algebra systems.


\end{abstract}

\section{Introduction}

The mathematical stage on which non-relativistic quantum mechanics is usually set is a Hilbert space. The state of a system is represented by a vector in this Hilbert space\footnote{Or a density operator. The framework described here is suitable for describing any operator, including density operators. I will not explicitly consider the density operator in this paper.}, physical observables are represented by linear self-adjoint operators, and the time evolution is dictated, through Schr\"odinger's equation, by the Hamiltonian operator.

It is often useful to pick a particular concrete example of the Hilbert space in which to solve problems. For instance, introductory quantum mechanics problems are usually solved explicitly in the space $L^2(\mathbb{R}^N)$, the space of square-integrable complex-valued functions --- so-called ``wavefunctions''. Problems in finite Hilbert spaces are often solved in the concrete space $\mathbb{C}^N$, the space of column vectors of complex numbers --- consider, for instance, problems featuring the familiar Pauli spin matrices.

It is well-known, though, that a more abstract approach to quantum mechanics can be very fruitful \cite{Isham}. Advanced calculations are often carried out without any reference to wavefunctions, or spin-matrices. It is not always necessary to specify a particular concrete Hilbert space, but rather the calculation proceeds by denoting vectors in the abstract space and working with them directly. I will refer to this abstract approach, which does not make essential reference to particular vector spaces, as the invariant formalism. These calculations are often carried out in a notation invented by Dirac which, as we shall see later, has had a significant impact on efforts to automate them.

Computer techniques have found many applications in quantum mechanics problems \cite{Feagin}. There exist a number of very sophisticated computer algebra packages \cite{Mathematica, Maple}. Almost any integral, sum, matrix calculation, or set of differential equations can be entered into one of these packages and automatically be solved or simplified. Numerical integration of differential equations permits the solution of problems depending on continuous parameters, including time-evolution problems. Matrix techniques can be used to solve problems with finite state-spaces \cite{qlib}. Computer algebra techniques have been used to automate the derivation of otherwise prohibitively complex expressions \cite{Hirata}.

However, despite the above developments, support for calculations in the invariant formalism is notably missing. A number of systems have been built that attempt to support these calculations \cite{Feagin, Maple, Harris, QDensity, Gomez}. But none of these correctly model all possible calculations in the invariant formalism, especially when tensor-product spaces are introduced. I will argue that this is a result of trying to implement Dirac's notation for the invariant formalism directly, rather than modelling the underlying mathematical structure.

In the sequel I will first consider in more detail the problems with basing a system for invariant formalism calculation directly on Dirac's notation. I will then present an alternative approach, as suggested above, founded on the underlying mathematical structure of quantum mechanics. I will provide a self-contained summary of this mathematical structure. I will then go on to characterise all possible algebraic expressions in this structure with a set of terms. After this, I will develop a set of rewrite rules that correspond to legal manipulations of the algebraic expressions. After a short example I will extend the system to Hilbert spaces with a tensor-product structure and present a fuller example.

\section{Dirac's notation}

Dirac developed a notation for calculations in the invariant formalism that is both powerful and easy to use \cite{Dirac}. His insight was to make the notation purposefully ambiguous. Most significantly, two types of symbol are introduced, bras and kets, that can be used to represent a variety of different mathematical objects. Alone these kets and bras represent state vectors and their duals respectively. In combination they can represent inner products, operators, and tensor products. The power of the notation lies in the fact that the syntactic ambiguity reflects, at least in most cases, some kind of equality between the denoted mathematical objects. Roughly speaking, if a combination of bras and kets `looks like' it could represent one of the above objects, it probably can. This flexibility and ambiguity makes the notation very comfortable to work with in a practical sense, reducing the number of tedious, mechanical steps in a calculation. Dirac's notation has deservedly found widespread acceptance.

Consider the following example, the application of a projection operator to a state in the Hilbert space $H$,
\begin{equation*}
\ket{\Psi} = P_{\ket{\phi}} \ket{\alpha}\ .
\end{equation*}
This can be written explicitly in Dirac's notation as
\begin{equation}
\label{dirac1}
\ket{\Psi} = (\ket{\phi} \bra{\phi}) \ket{\alpha}\ ,
\end{equation}
where I have bracketed the projection operator to emphasise that this expression represents an operator that maps $H \rightarrow H$, acting on an element of $H$. However, this expression can also be interpreted as
\begin{equation}
\label{dirac2}
\ket{\Psi} = \ket{\phi} (\braket{\phi | \alpha})\ ,
\end{equation}
where the expression is now to be read as an element of $H$ multiplied by a complex number, the inner-product of two vectors. The beauty of Dirac's notation is that both of the above expressions are extensionally equal. Conventionally, the expression would not be bracketed, giving the reader the flexibility to read the expression both ways (\ref{dirac1} and \ref{dirac2}).

The scalar multiplication in (\ref{dirac2}) is commutative and so the expression can be rewritten as
\begin{equation*}
\ket{\Psi} = \braket{\phi | \alpha} \ket{\phi}\ ,
\end{equation*}
It is, however, not acceptable to rearrange equation \ref{dirac1} to give
\begin{equation*}
\ket{\Psi} \neq   \braket{\phi | \phi} \ket{\alpha}\\ .
\end{equation*}

Although any practising quantum-physicist would immediately recognise the above manipulation as invalid, most would be unable to easily provide a formal specification of which manipulations are allowed and which are not. Working in Dirac's notation is a skill that is learnt and the rules are complex. In the author's experience, many physics students struggle to understand the subtleties of manipulation of complex Dirac expressions, particularly where tensor product spaces and operators are involved.

Formally, Dirac's notation represents five separate mathematical operations with the same syntactic form: juxtaposition. The operations of scalar multiplication, taking the inner product, operator application, operator composition, and forming the tensor product\footnote{Sometimes the tensor product is explicitly indicated in particularly ambiguous, or didactic, contexts.} are all represented by butting bras and kets up against one another. The meaning of a bra or a ket in a Dirac expression depends heavily on its context, and often there can be more than one context that is applicable for a given expression. Mechanical rules for re-arranging the symbols are thus difficult to provide because whether it is admissible to manipulate a symbol depends on the symbol's context. To successfully work in Dirac's notation one must understand the meaning of the expressions, as it is these meanings that provide the context. It is this point that makes a computer implementation of Dirac's notation very difficult. Computer algebra systems have no notion of meaning --- they are automata that perform mechanistic structural re-arrangement of expressions.

Previous software packages have typically tried to capture the intricacies of Dirac's notation by treating expressions as sequences of bras and kets glued together by generalised `multiplication' operators. Some control is usually provided over the commutativity properties of the bras and kets with respect to these generalised multiplications, with varying degrees of sophistication. These commutativity properties are then tuned to match the complex manipulation rules of Dirac's notation. This approach is sufficient to model simple manipulations of Dirac expressions. However, as these systems are not capturing the full context associated with these expressions, they are bound to be limited in their scope. To fully capture this notion of context I argue that one must model the full underlying mathematical structure. In this paper then, rather than attempt to implement a computer algebra system for Dirac's notation directly, I will describe a computer algebra system for the underlying mathematical structure. If desired, it is reasonably straightforward to add Dirac's notation as an ambiguous input/output representation.

\section{Quantum algebra}
\label{algebra}

In this section I describe the mathematical formalism behind (non-relativistic) quantum mechanics. The treatment will be brief and serves to provide a self-contained definition to refer to. The reader is directed towards \cite{Isham} for more details. In the rest of the paper I will loosely describe this mathematical structure as the ``quantum algebra''. For simplicity of presentation I shall ignore the tensor product for the moment. I will remove this simplification in section \ref{tensor}.

The fundamental elements of the algebra are (complex) Hilbert space vectors. I will denote vectors by an overhead arrow e.g. $\vec{\psi}$\ . Complex linear combinations of vectors are also vectors
\begin{equation*}
\label{vec}
\vec{\phi} = 3 \vec{\chi} + 4 i \vec{\psi}\ ,
\end{equation*}
where, as usual, the addition operation is associative and commutative.

The space is equipped with an inner product, which is a sesquilinear operator that maps two vectors into a complex number. The inner product is denoted, following the usual mathematicians' convention, as a round-bracketed pair of vectors, separated by a comma. Sesquilinearity means the following properties hold
\begin{align}
\label{ip}
(\vec{\phi}, a\,\vec{\psi} + b\,\vec{\chi}) &= a (\vec{\phi}, \vec{\psi}) + b (\vec{\phi}, \vec{\chi})\ ,\\
(a\,\vec{\phi} + b\,\vec{\psi}, \vec{\chi}) &= a^* (\vec{\phi}, \vec{\chi}) + b^* (\vec{\psi}, \vec{\chi})\ ,
\end{align}
where I have followed the physicists' convention of making the product conjugate-linear in its first argument.

Operators are maps from a Hilbert space into itself\footnote{Loosely speaking. There is some subtlety to do with domains that is rarely important.}. They will be denoted by overhead hats, as usual. Operators are linear maps, in the sense that
\begin{equation}
\label{opap1}
\hat{O}(a\,\vec{\phi} + b\,\vec{\psi}) = a\,\hat{O}(\vec{\phi}) + b\,\hat{O}(\vec{\psi})\ .
\end{equation}
Operators are also elements of a complex vector space, so complex linear combinations of operators are themselves operators,
\begin{equation*}
\label{op}
\hat{O} = \hat{P} + 3 i \hat{Q}\ .
\end{equation*}
As with vectors, this addition operator is associative and commutative. Operator application is linear with respect to the vector space structure of operators,
\begin{equation}
\label{opap2}
(a\, \hat{O} + b\, \hat{P})(\vec{\phi}) = a\,\hat{O}(\vec{\phi}) + b\,\hat{P}(\vec{\phi})\ .
\end{equation}
Operators can be composed to form new operators\footnote{This is sometimes misleadingly referred to as multiplication of operators.}. I will represent composition by a dot,
\begin{equation*}
\hat{O} = \hat{P} \cdot \hat{Q}\ .
\end{equation*}
The action of a composite operator on a state is defined by
\begin{equation}
\label{opcomp}
(\hat{P} \cdot \hat{Q}) \vec{\psi} = \hat{P}(\hat{Q} (\vec{\psi}))\ .
\end{equation}

The final objects to deal with are Dirac's bras. Dirac introduces the dual space to the Hilbert space of kets, the space of bras. This introduction is unnecessary and some authors have argued that it creates more problems than it solves \cite{Gieres}. I will take an alternative approach. It is sufficient to introduce a projection operator, parameterised by two vectors, with definition
\begin{equation}
\label{proj}
\hat{M}[\vec{\psi}, \vec{\phi}](\vec{\theta}) = ( \vec{\phi}, \vec{\theta} )\,\vec{\psi}\ .
\end{equation}
It is straightforward to show that this does indeed satisfy the linearity requirement for an operator. This approach has the advantage that rules for operators will apply to the projector, minimising the amount of duplication. The operator $\hat{M}$ is equivalent to the Dirac construction
\begin{equation*}
\hat{M}[\vec{\psi}, \vec{\phi}] \equiv \ket{\psi} \bra{\phi}\ .
\end{equation*}
All expressions featuring bras in Dirac's notation can either be rewritten using the above projection operators, or else their dual can.\footnote{This is not strictly true, as one could imagine a space that is constructed as the tensor product of a ket-space and a bra-space $\ket{a} \otimes \bra{b}$. For all practical purposes, though, this could be replaced with the equivalent ket-ket space $\ket{a} \otimes \ket{b}$, and the projection operators are once again sufficient.}

Bases are not a fundamental, nor necessary, ingredient of the algebra. They are however often essential to make progress in a calculation. We will consider bases again in section \ref{teleportation}.

\section{Term structure}
\label{terms}

I have described above the mathematical structure that underlies quantum mechanical calculations. Our aim in this paper is to develop a framework for automating calculations in this algebra. The computer is a mechanistic device, and as such we need to provide it with a set of mechanical rules for valid, structural transformations of quantum algebra expressions. The first step in developing these rules is to characterise the structure of possible expressions in the quantum algebra. To effect this characterisation we need to consider a second, very different, mathematical construction: a set of terms.

The notion of a term comes from the field of universal algebra \cite{traat}. Informally, a term looks like a nested set of function-calls i.e. $f(x, g(y))$ or $h()$. Formally, a set of terms is defined by a signature, $\Sigma$, and a set of variables, $X$. The signature is a set of function symbols ($f$, $g$, and $h$ in the above examples) and their \emph{arities}. The arity of a function symbol is a specification of how many, and what \emph{sort} of, arguments it takes. The sort of the argument is required because there are distinct types of mathematical objects in our quantum algebra: vectors, operators and scalars. It does not, for example, make sense to take the inner product of a vector and an operator. In the literature of term-rewriting and universal algebra a set of terms with sorted\footnote{This confusing use of the word sorted is, unfortunately, established in the literature. It should be read as ``having a sort'', not the more usual ``put in order''.} arities, is known as a many-sorted term system. The variables $X$ are simply a set of symbols, distinct from the function symbols in the signature. They can be used to stand for any valid term in the definition of a rule (see section \ref{rules}). Terms are defined inductively, as either a variable, or a function symbol applied to other valid terms of a suitable sort.

To model the structure of expressions in the quantum algebra I introduce the following signature. First, there are function symbols for combining scalars, with their associated arities
\begin{align*}
\mathtt{conjugate} && scalar &\rightarrow scalar\ ,\\
\mathtt{plusS} && scalar \times scalar &\rightarrow scalar\ ,\\
\mathtt{timesS} && scalar \times scalar &\rightarrow scalar\ .
\end{align*}
where the arity of $\mathtt{plusS}$, for instance, should be understood as indicating that the function symbol $\mathtt{plusS}$ requires two scalar arguments, and is itself of scalar sort. This is a minimal set of operations on scalars. Any realistic computer algebra system will have a much larger set of scalar representatives (integrals, sums, fractions and the like). As computer algebra techniques for scalars are well-established I will not consider them further, and keep only this minimal set of operations.

Next, there are function symbols that represent linear combinations of vectors and operators
\begin{align*}
\mathtt{plusV} && vector \times vector &\rightarrow vector\ ,\\
\mathtt{timesV} && scalar \times vector &\rightarrow vector\ ,\\
\mathtt{plusO} && operator \times operator &\rightarrow operator\ ,\\
\mathtt{timesO} && scalar \times operator &\rightarrow operator\ .
\end{align*}

These are followed by function symbols to represent the operations of taking the inner product, applying an operator, and composing two operators
\begin{align*}
\mathtt{ip} && vector \times vector &\rightarrow scalar\ ,\\
\mathtt{apply} && operator \times vector &\rightarrow vector\ ,\\
\mathtt{compose} && operator \times operator &\rightarrow operator\ .
\end{align*}

Finally, I introduce a function symbol to represent the projection operators (equation \ref{proj}) that will take the place of Dirac's kets
\begin{align*}
\mathtt{projector} && vector \times vector &\rightarrow operator\ .
\end{align*}

In addition to the above function symbols I shall assume that there is a ready supply of constant (unary) function symbols of all sorts for representing constant vectors, scalars, and operators. I will introduce variables in section \ref{rules}.

Let us consider an example. The Dirac expression
\begin{equation*}
(2 \hat{p} + 5 \hat{q})(3 \ket{\psi} + \braket{\alpha | \beta} \ket{\phi})\ ,
\end{equation*}
is represented by the term
\begin{align*}
\mathtt{apply}(&\\
		&\mathtt{plusO}(\ \mathtt{timesO}(2, \hat{p}),\ \mathtt{timesO}(5, \hat{q})\ ),\\
		&\mathtt{plusV}(\ \mathtt{timesV}(3, \vec{\psi}),\  \mathtt{timesV}(\ \mathtt{ip}(\vec{\alpha},\vec{\beta}),\ \vec{\phi}\ )\ )\\
)&
\end{align*}
where $2, 3, 5, \hat{p}, \hat{q}, \vec{\alpha}, \vec{\beta}, \vec{\phi}$, and $ \vec{\psi}$ are constant symbols of $scalar$, $operator$ and $vector$ sorts respectively.

It might seem as though I've just invented a new and rather clunky notation for expressions in the quantum algebra. But the significance is deeper than that: the term specification above gives a formal definition of \emph{all} possible expressions in the quantum algebra and what sort of expression they are. The beauty of the term representation is in its regularity and lack of ambiguity.

\section{Rewrite rules}
\label{rules}

In section \ref{algebra} the mathematical structure of quantum mechanics calculations in the invariant formalism was outlined. Then, in section \ref{terms}, a second mathematical structure was introduced, one that describes every possible \emph{expression} that can be written in the algebra of section \ref{algebra}. In this section I will make a link between these two structures by describing which term-expressions correspond to things that, in the quantum algebra, have the same \emph{value}. This is the essence of algebraic manipulation: changing the form of an expression without changing its value.

The mathematical tool that I will use to make the link is what is known as a term-rewriting system (TRS). A TRS is a set of terms, as defined above, and a set of rules that transform one term into another. A rule is applied by taking a term, or part of a term, that matches the left hand side of the rule and replacing it with the right hand side. Rules can contain variables which can represent any term of the appropriate sort. Readers seeking formal definitions of a rule, matching, and rule application will find them in \cite{traat}.

The rules for the quantum algebra are as follows. The rules are written with left-right arrows to indicate that, as described below, they can be used in both directions. I will assume that suitable rules for the pure scalar operations, $\mathtt{conjugate}$, $\mathtt{plusS}$, and $\mathtt{timesS}$, will be built in to any computer algebra system and will not present them here. The first three rules rewrite linear combinations of vectors
\begin{align}
\label{expandRightV}
\mathtt{timesV}(\ a,\ \mathtt{plusV}(v_1, v_2)\ ) &\leftrightarrow \mathtt{plusV}(\ \mathtt{timesV}(a, v_1),\ \mathtt{timesV}(a, v_2)\ )\ ,\\
\label{expandLeftV}
\mathtt{timesV}(\ \mathtt{plusS}(a_1, a_2),\ v_1\ ) &\leftrightarrow \mathtt{plusV}(\ \mathtt{timesV}(a_1, v),\ \mathtt{timesV}(a_2, v)\ )\ ,\\
\label{multiplyLeftV}
\mathtt{timesV}(\ a_1,\ \mathtt{timesV}(a_2, v)\ ) &\leftrightarrow \mathtt{timesV}(\ \mathtt{timesS}(a_1, a_2),\ v\ )\ .
\end{align}
where here, and in what follows, symbols beginning $a, v$ and $o$ are variables (members of $X$) that can stand for any term of scalar, vector, and operator sort respectively.

The next three rules are the equivalent of the above for operators
\begin{align}
\label{expandRightO}
\mathtt{timesO}(\ a,\ \mathtt{plusO}(o_1, o_2)\ ) &\leftrightarrow \mathtt{plusO}(\ \mathtt{timesO}(a, o_1),\ \mathtt{timesO}(a, o_2)\ )\ ,\\
\label{expandLeftO}
\mathtt{timesO}(\ \mathtt{plusS}(a_1, a_2),\ o\ ) &\leftrightarrow \mathtt{plusO}(\ \mathtt{timesO}(a_1, o),\ \mathtt{timesO}(a_2, o)\ )\ ,\\
\label{multiplyLeftO}
\mathtt{timesO}(\ a_1,\ \mathtt{timesO}(a_2, o)\ ) &\leftrightarrow \mathtt{timesO}(\ \mathtt{timesS}(a_1, a_2),\ o\ )\ .
\end{align}

The sesquilinearity of the inner product (\ref{ip}) is captured with the next set of rules
\begin{align}
\label{expandRightIP}
\mathtt{ip}(\ v_1,\ \mathtt{plusV}(v_2, v_3)\ ) &\leftrightarrow \mathtt{plusS}(\ \mathtt{ip}( v_1, v_2),\ \mathtt{ip}( v_1, v_3\ )\ )\ ,\\ 
\label{expandLeftIP}
\mathtt{ip}(\ \mathtt{plusV}(v_1, v_2),\ v_3\ ) &\leftrightarrow \mathtt{plusS}(\ \mathtt{ip}( v_1, v_3),\ \mathtt{ip}( v_2, v_3)\ )\ ,\\
\label{multiplyRightIP}
\mathtt{ip}(\ v_1,\ \mathtt{timesV}(a, v_2)\ ) &\leftrightarrow \mathtt{timesS}(\ a,\ \mathtt{ip}( v_1, v_2)\ )\ ,\\
\label{multiplyLefttIP}
\mathtt{ip}(\ \mathtt{timesV}(a, v_1),\ v_2\ ) &\leftrightarrow \mathtt{timesS}(\ \mathtt{conjugate}(a),\ \mathtt{ip}( v_1, v_2)\ )\ .
\end{align}

The following set of rules deal with the bilinearity of operator application (\ref{opap1} and \ref{opap2}) and the action of composite operators (\ref{opcomp})
\begin{align}
\label{expandRightApply}
\mathtt{apply}(\ o,\ \mathtt{plusV}(v_1, v_2)\ ) &\leftrightarrow \mathtt{plusV}(\ \mathtt{apply}(o, v_1),\  \mathtt{apply}(o, v_2)\ )\ ,\\
\label{multiplyRightApply}
\mathtt{apply}(\ o,\ \mathtt{timesV}(a, v)\ ) &\leftrightarrow \mathtt{timesV}(\ a,\ \mathtt{apply}(o, v)\ )\ ,\\
\label{expandLeftApply}
\mathtt{apply}(\ \mathtt{plusO}(o_1, o_2),\ v\ ) &\leftrightarrow \mathtt{plusV}(\ \mathtt{apply}(o_1, v),\  \mathtt{apply}(o_2, v)\ )\ ,\\
\label{multiplyLeftApply}
\mathtt{apply}(\ \mathtt{timesO}(a, o),\ v\ ) &\leftrightarrow \mathtt{timesV}(\ a,\ \mathtt{apply}(o, v)\ )\ ,\\
\label{expandCompose}
\mathtt{apply}(\ \mathtt{compose}(o_1,o_2),\ v\ ) &\leftrightarrow \mathtt{apply}(\ o_1,\ \mathtt{apply}(o_2, v)\ )\ .
\end{align}

The next rule describes the operation of the projection operators (\ref{proj})
\begin{align}
\label{applyProjector}
\mathtt{apply}(\ \mathtt{projector}(v_1, v_2),\ v_3\ ) &\leftrightarrow \mathtt{timesV}(\ \mathtt{ip}(v_2, v_3),\ v_1\ )\ .
\end{align}

Finally, the following rules capture the associative and commutative nature of vector and operator addition
\begin{align}
\label{commuteV}
\mathtt{plusV}(v_1, v_2) &\leftrightarrow \mathtt{plusV}(v_2, v_1)\ ,\\
\label{assocV}
\mathtt{plusV}(\ v_1,\ \mathtt{plusV}(v_2, v_3)\ ) &\leftrightarrow \mathtt{plusV}(\ \mathtt{plusV}(v_1, v_2),\ v_3\ )\ ,\\
\label{commuteO}
\mathtt{plusO}(o_1, o_2) &\leftrightarrow \mathtt{plusO}(o_2, o_1)\ ,\\
\label{assocO}
\mathtt{plusO}(\ o_1,\ \mathtt{plusO}(o_2, o_3)\ ) &\leftrightarrow \mathtt{plusO}(\ \mathtt{plusO}(o_1, o_2),\ o_3\ )\ .
\end{align}
It is usual to avoid adding explicit rules such as these to capture commutativity and associativity, as they lend the TRS some undesirable properties. Rather, rule matching is usually extended to include matching modulo associativity or commutativity. For our purposes, though, this will add unnecessary complication and, for now, adding these associative and commutative rules will have no ill effect.

The rules allow us to manipulate expressions without changing their value. Any expression can be transformed into any other with the same value by successive application of these rules, either in the forward of reverse direction. Formally, the transitive reflexive closure of these rules generates an equivalence relation on the set of terms. The equivalence classes of this relation are precisely the sets of terms that correspond to expressions in the quantum algebra that have the same value.

\section{Simple example}

Let us consider an example to demonstrate the action of the TRS. We will expand the application of a projection operator to a state
\begin{equation*}
\ket{\alpha}\bra{\alpha}(\frac{1}{\sqrt{2}}(\ket{\beta} - \ket{\gamma}))\ .
\end{equation*}

To describe the application of rules from the TRS to parts of an expression we need to define \emph{position} within a term. Each part of a term, apart from the term itself, is the argument to a function symbol. Thus, any part of a term can be specified by saying which argument of the enclosing function symbol it is, and which argument the enclosing function symbol is of its enclosing symbol, and so on, until the root of the term is reached. This list of integers is the position. Conventionally the list is ordered to start from the root. The term itself is given the position $\epsilon$.

\begin{sidewaystable}[!p]
\centering
\setlength{\tabcolsep}{1pt}

\begin{tabular}{|l|l|c|c|}

\hline
Term & Dirac-form & Rule & Position \\
\hline
$\mathtt{apply}(\ \mathtt{projector}(\ \vec{\alpha },\ \vec{\alpha }\ ),\ \mathtt{timesV}(\ \frac{1}{\sqrt{2}},\ \mathtt{plusV}(\ \vec{\beta },\ \mathtt{timesV}(\ -1,\ \vec{\gamma }\ )\ )\ )\ )$&
$\ket{\alpha}\bra{\alpha}(\frac{1}{\sqrt{2}}(\ket{\beta} + (- \ket{\gamma})))$&
\ref{multiplyRightApply}\ $\rightarrow$&
$\epsilon$\\
&&&\\
$\mathtt{timesV}(\ \frac{1}{\sqrt{2}},\ \mathtt{apply}(\ \mathtt{projector}(\ \vec{\alpha },\ \vec{\alpha }\ ),\ \mathtt{plusV}(\ \vec{\beta },\ \mathtt{timesV}(\ -1,\ \vec{\gamma }\ )\ )\ )\ )$&
$\frac{1}{\sqrt{2}} \ket{\alpha}\bra{\alpha} (\ket{\beta} + (- \ket{\gamma}))$&
\ref{expandRightApply}\ $\rightarrow$&
$2$\\
&&&\\
$\mathtt{timesV}(\ \frac{1}{\sqrt{2}},\ \mathtt{plusV}(\ \mathtt{apply}(\ \mathtt{projector}(\ \vec{\alpha },\ \vec{\alpha }\ ),\ \vec{\beta }\ ),\ \mathtt{apply}(\ \mathtt{projector}(\ \vec{\alpha },\ \vec{\alpha }\ ),\ \mathtt{timesV}(\ -1,\ \vec{\gamma }\ )\ )\ )\ )$&
$\frac{1}{\sqrt{2}} (\ket{\alpha}\bra{\alpha}(\ket{\beta}) + \ket{\alpha}\bra{\alpha}(-\ket{\gamma}))$&
\ref{multiplyRightApply}\ $\rightarrow$&
$2,2$\\
&&&\\
$\mathtt{timesV}(\ \frac{1}{\sqrt{2}},\ \mathtt{plusV}(\ \mathtt{apply}(\ \mathtt{projector}(\ \vec{\alpha },\ \vec{\alpha }\ ),\ \vec{\beta }\ ),\ \mathtt{timesV}(\ -1,\ \mathtt{apply}(\ \mathtt{projector}(\ \vec{\alpha },\ \vec{\alpha }\ ),\ \vec{\gamma }\ )\ )\ )\ )$&
$\frac{1}{\sqrt{2}} (\ket{\alpha}\bra{\alpha}(\ket{\beta}) - \ket{\alpha}\bra{\alpha}(\ket{\gamma}))$&
\ref{applyProjector}\ $\rightarrow$&
$2,1$\\
&&&\\
$\mathtt{timesV}(\ \frac{1}{\sqrt{2}},\ \mathtt{plusV}(\ \mathtt{timesV}(\ \mathtt{ip}(\ \vec{\alpha },\ \vec{\beta }\ ),\ \vec{\alpha }\ ),\ \mathtt{timesV}(\ -1,\ \mathtt{apply}(\ \mathtt{projector}(\ \vec{\alpha },\ \vec{\alpha }\ ),\ \vec{\gamma }\ )\ )\ )\ )$&
$\frac{1}{\sqrt{2}} (\braket{\alpha,\beta}\ket{\alpha}  - \ket{\alpha}\bra{\alpha}(\ket{\gamma}))$&
\ref{applyProjector}\ $\rightarrow$&
$2,2,2$\\
&&&\\
$\mathtt{timesV}(\ \frac{1}{\sqrt{2}},\ \mathtt{plusV}(\ \mathtt{timesV}(\ \mathtt{ip}(\ \vec{\alpha },\ \vec{\beta }\ ),\ \vec{\alpha }\ ),\ \mathtt{timesV}(\ -1,\ \mathtt{timesV}(\ \mathtt{ip}(\ \vec{\alpha },\ \vec{\gamma }\ ),\ \vec{\alpha }\ )\ )\ )\ )$&
$\frac{1}{\sqrt{2}} (\braket{\alpha,\beta}\ket{\alpha} - \braket{\alpha,\gamma} \ket{\alpha})$&
\ref{multiplyLeftV}\ $\rightarrow$&
$2,2$\\
&&&\\
$\mathtt{timesV}(\ \frac{1}{\sqrt{2}},\ \mathtt{plusV}(\ \mathtt{timesV}(\ \mathtt{ip}(\ \vec{\alpha },\ \vec{\beta }\ ),\ \vec{\alpha }\ ),\ \mathtt{timesV}(\ \mathtt{timesS}(\ -1,\ \mathtt{ip}(\ \vec{\alpha },\ \vec{\gamma }\ )\ ),\ \vec{\alpha }\ )\ )\ )$&
$\frac{1}{\sqrt{2}} (\braket{\alpha,\beta}\ket{\alpha} + (- \braket{\alpha,\gamma}) \ket{\alpha})$&
\ref{expandLeftV}\ $\leftarrow$&
$2$\\
&&&\\
$\mathtt{timesV}(\ \frac{1}{\sqrt{2}},\ \mathtt{timesV}(\ \mathtt{plusS}(\ \mathtt{ip}(\ \vec{\alpha },\ \vec{\beta }\ ),\ \mathtt{timesS}(\ -1,\ \mathtt{ip}(\ \vec{\alpha },\ \vec{\gamma }\ )\ )\ ),\ \vec{\alpha }\ )\ )$&
$\frac{1}{\sqrt{2}} ((\braket{\alpha,\beta} - \braket{\alpha,\gamma}) \ket{\alpha})$&
\ref{multiplyLeftV}\ $\rightarrow$&
$\epsilon$\\
&&&\\
$\mathtt{timesV}(\ \mathtt{timesS}(\ \frac{1}{\sqrt{2}},\ \mathtt{plusS}(\ \mathtt{ip}(\ \vec{\alpha },\ \vec{\beta }\ ),\ \mathtt{timesS}(\ -1,\ \mathtt{ip}(\ \vec{\alpha },\ \vec{\gamma }\ )\ )\ )\ ),\ \vec{\alpha }\ )$&
$(\frac{1}{\sqrt{2}} (\braket{\alpha,\beta} - \braket{\alpha,\gamma})) \ket{\alpha}$&
---&
---\\

\hline
\end{tabular}
\caption{Application of a projection operator. The first column shows the term as successive rewrite rules are applied to it. The second column shows the equivalent Dirac-form. More bracketing than would be usual has been used to better reflect the structure of the term. The third column shows the rule that has been applied and the direction in which it was used. The final column shows the position at which the rule was applied.}
\label{example1}
\end{sidewaystable}

Table \ref{example1} shows step-by-step how the term evolves. First the scalar pre-factor is brought out of the operator application. The second step uses the linearity of the operator to split the expression into a sum of two vectors. After removing another scalar factor from an operator application in the third step, the projection operators are applied. Finally, the expression is rearranged into a conventional form by manipulating the scalar mulitplying the vector. Note that if the Dirac-form expressions were written conventionally, without the extra bracketing, a number of expressions at successive steps would be identical. 

This example may seem facile, but it is not. The important point is that at each step the expression had an unambiguous representation and each step involved applying a particular rule to a well-defined position. It is this rigorously defined mechanical precision that is essential to a well-functioning computer algebra system.

\section{Extension to tensor product spaces}
\label{tensor}

Thus far we have only dealt with the quantum mechanics of isolated, single systems. To accommodate the description of multipartite systems we need to remove the simplification made in section \ref{algebra} and introduce the tensor product.

If the states of two systems are represented by vectors in the Hilbert spaces $\mathcal{H}_1$ and $\mathcal{H}_2$ respectively\footnote{Often these spaces will be, in a formal sense, the same space --- consider, for instance, two coupled qubits. We will nonetheless treat these spaces as distinct as they pertain to different physical systems.}, then states of the composite system are represented by vectors in the tensor product space $\mathcal{H}_1 \otimes \mathcal{H}_2$. It is this tensor product structure that is responsible for many interesting effects in quantum mechanics. In this section I consider how to extend the term-rewriting system to tensor product spaces.

\subsection{Quantum algebra}

The quantum algebra is extended as follows. The states will be labelled with a subscript that indicates the space e.g. $\vec{\psi}_1$\ is a vector in space $\mathcal{H}_1$.

The tensor product of two vectors will be indicated in the usual way,
\begin{equation*}
\vec{\psi}_1 \otimes \vec{\phi}_2\ .
\end{equation*}
The tensor product operator is associative and commutative. The product is linear in its two arguments
\begin{equation}
\label{tensorVLin}
\begin{split}
\vec{\phi}_{1} \otimes (a\,\vec{\psi}_{2} + b\,\vec{\theta}_2) &= a(\vec{\phi}_1 \otimes \vec{\psi}_2) + b(\vec{\phi}_1 \otimes \vec{\theta}_2)\ ,\\
(a\,\vec{\phi}_{1} + b\,\vec{\psi}_{1}) \otimes \vec{\theta}_2 &= a(\vec{\phi}_1 \otimes \vec{\theta}_2) + b(\vec{\psi}_1 \otimes \vec{\theta}_2)\ .
\end{split}
\end{equation}
It is necessary to extend the definition of the inner-product over tensor products of vectors,
\begin{equation}
\label{tip}
(\vec{\psi}_1 \otimes \vec{\phi}_2, \vec{\theta}_1 \otimes \vec{\beta}_2) = (\vec{\psi}_1, \vec{\theta}_1) (\vec{\phi}_2, \vec{\beta}_2)\ .
\end{equation}
Note that this inner product is only defined if the two vectors belong to the same space.

Operators are also labelled with a subscript to indicate the space they act on. Tensor products of operators can be defined, and as with vectors, the operator tensor product is associative, commutative, and linear in its arguments
\begin{equation}
\label{tensorOLin}
\begin{split}
\hat{p}_{1} \otimes (a\,\hat{q}_{2} + b\,\hat{r}_2) &= a(\hat{p}_1 \otimes \hat{q}_2) + b(\hat{p}_1 \otimes \hat{r}_2)\ ,\\
(a\,\hat{p}_{1} + b\,\hat{q}_{1}) \otimes \hat{r}_2 &= a(\hat{p}_1 \otimes \hat{r}_2) + b(\hat{q}_1 \otimes \hat{r}_2)\ .
\end{split}
\end{equation}
The action of a tensor product operator on a tensor product of vectors is defined by,
\begin{equation}
\label{tensorOp}
\hat{O}_1 \otimes \hat{P}_2 ( \vec{\psi}_1 \otimes \vec{\phi}_2 ) = \hat{O}_1 (\vec{\psi}_1) \otimes \hat{P}_2 ( \vec{\phi}_2 )\ .
\end{equation}
As with vectors, a subscript on an operator indicates which space it acts on\footnote{Note that only tensor product operators can act on tensor product states. Often when an operator in one space is tensored with the identity operator in another, the identity is omitted. That is not admissable here.}.

\subsection{Terms}
\label{tensorTerms}
Vectors, and operators can now belong to different Hilbert spaces. We embed this information in the term structure by parameterising the sorts. We will represent vectors and operators associated with the space $\mathcal{H}_1 \otimes \mathcal{H}_2 \otimes \cdots$ by terms with sorts
\begin{equation*}
\begin{split}
vector[\mathcal{H}_1 \otimes \mathcal{H}_2 \otimes \cdots]\ ,\\
operator[\mathcal{H}_1 \otimes \mathcal{H}_2 \otimes \cdots]\ .
\end{split}
\end{equation*}
The term structure of section \ref{terms} is extended in the obvious way using the new parameterised sorts. Two additional function symbols are needed to express tensor products of vectors and operators respectively
\begin{align*}
\mathtt{tensorV} && vector[\mathcal{H}_1] \times vector[\mathcal{H}_2] &\rightarrow vector[\mathcal{H}_1 \otimes \mathcal{H}_2]\ , \\
\mathtt{tensorO} && operator[\mathcal{H}_1] \times operator[\mathcal{H}_2] &\rightarrow operator[\mathcal{H}_1 \otimes \mathcal{H}_2]\ .
\end{align*}
The use of the tensor product notation inside the parameterised sort is more than just a notational convenience. It is meant to be taken as implying that there are isomorphisms from the semigroup of Hilbert spaces to the sorts of vectors and operators, with respect to the tensor product operation. For example, consider two tensor product expressions of vectors from four vector spaces
\begin{align*}
\mathtt{tensorV}( v_1, v_2 )&\ , \textrm{and} \\
\mathtt{tensorV}( v_3, v_4 )& 
\end{align*}
where,
\begin{align*}
\textbf{sort}( v_1 ) &= vector[\mathcal{H}_1 \otimes \mathcal{H}_2]\ ,\\
\textbf{sort}( v_2 ) &= vector[\mathcal{H}_3]\ ,\\
\textbf{sort}( v_3 ) &= vector[\mathcal{H}_1]\ ,\\
\textbf{sort}( v_4 ) &= vector[\mathcal{H}_2 \otimes \mathcal{H}_3]\ .
\end{align*}
I have introduced the operator $\textbf{sort}$ which reveals the sort of a term. The isomorphism requires that the structure of the tensor product operation is respected by the sort structure, meaning that in this case the following must hold
\begin{equation*}
\textbf{sort}\big( \mathtt{tensorV}( v_1, v_2) \big) =  
\textbf{sort}\big( \mathtt{tensorV}( v_3, v_4) \big)\ .
\end{equation*}

\subsection{Rewrite rules}

We need to extend the rewrite rules to work with tensor products. The first set of additional rules capture the linearity of the tensor product operation with respect to vectors (\ref{tensorVLin})
\begin{align}
\mathtt{tensorV}( v_1, \mathtt{plusV}(v_2, v_3) ) &\leftrightarrow \mathtt{plusV}(\mathtt{tensorV}( v_1, v_2), \mathtt{tensorV}( v_1, v_3))\ ,\\ 
\mathtt{tensorV}( \mathtt{plusV}(v_1, v_2), v_3 ) &\leftrightarrow \mathtt{plusV}(\mathtt{tensorV}( v_1, v_3), \mathtt{tensorV}( v_2, v_3))\ ,\\
\mathtt{tensorV}( \mathtt{timesV}(a, v_1), v_2 ) &\leftrightarrow \mathtt{timesV}(a, \mathtt{tensorV}( v_1, v_2))\ ,\\
\mathtt{tensorV}( v_1, \mathtt{timesV}(a, v_2) ) &\leftrightarrow \mathtt{timesV}(a, \mathtt{tensorV}( v_1, v_2))\ .
\end{align}
An equivalent set of rules are needed for tensor products of operators (\ref{tensorOLin})
\begin{align}
\mathtt{tensorO}( o_1, \mathtt{plusO}(o_2, o_3) ) &\leftrightarrow \mathtt{plusO}(\mathtt{tensorO}( o_1, o_2), \mathtt{tensorO}( o_1, o_3))\ ,\\ 
\mathtt{tensorO}( \mathtt{plusO}(o_1, o_2), o_3 ) &\leftrightarrow \mathtt{plusO}(\mathtt{tensorO}( o_1, o_3), \mathtt{tensorO}( o_2, o_3))\ ,\\
\mathtt{tensorO}( \mathtt{timesO}(a, o_1), o_2 ) &\leftrightarrow \mathtt{timesO}(a, \mathtt{tensorO}( o_1, o_2))\ ,\\
\mathtt{tensorO}( o_1, \mathtt{timesO}(a, o_2) ) &\leftrightarrow \mathtt{timesO}(a, \mathtt{tensorO}( o_1, o_2))\ .
\end{align}
Once again, explicit rules are added to represent the associativity and commutativity of the vector and operator tensor products
\begin{align}
\label{commuteTV}
\mathtt{tensorV}(v_1, v_2) &\leftrightarrow \mathtt{tensorV}(v_2, v_1)\ ,\\
\label{assocTV}
\mathtt{tensorV}(\ v_1,\ \mathtt{tensorV}(v_2, v_3)\ ) &\leftrightarrow \mathtt{tensorV}(\ \mathtt{tensorV}(v_1, v_2),\ v_3\ )\ ,\\
\label{commuteTO}
\mathtt{tensorO}(o_1, o_2) &\leftrightarrow \mathtt{tensorO}(o_2, o_1)\ ,\\
\label{assocTO}
\mathtt{tensorO}(\ o_1,\ \mathtt{tensorO}(o_2, o_3)\ ) &\leftrightarrow \mathtt{tensorO}(\ \mathtt{tensorO}(o_1, o_2),\ o_3\ )\ .
\end{align}
Inner products of tensor products (\ref{tip}) are defined by the rule
\begin{align}
\mathtt{ip}(\mathtt{tensorV}(v_1,v_2), \mathtt{tensorV}(v_3,v_4)) &\leftrightarrow \mathtt{timesS}(\mathtt{ip}(v_1, v_3),  \mathtt{ip}(v_2, v_4))\ .
\end{align}
And, finally, the operation of tensor product operators on tensor product states (\ref{tensorOp})
\begin{align}
\mathtt{apply}(\mathtt{tensorO}(o_1, o_2), \mathtt{tensorV}(v_1, v_2)) &\leftrightarrow \mathtt{tensorV}(\mathtt{apply}(o_1, v_1), \mathtt{apply}(o_2, v_2))\ .
\end{align}

\section{Example: quantum teleportation}
\label{teleportation}
Quantum teleportation is a well-known illustration of the behaviour of entangled systems \cite{Chuang}. Alice and Bob share a pair of qubits that they have previously entangled into a Bell-state
\begin{equation}
\label{BellState}
\ket{\theta} = \frac{1}{\sqrt{2}} ( \ket{0}_{a}\ket{0}_b + \ket{1}_{a}\ket{1}_b )\ .
\end{equation}
where $a$ is the Hilbert space of Alice's qubit and $b$ is the space of Bob's qubit. All of the states are written in the so-called computational basis. Alice wishes to transfer the state of a second qubit, in space $a_2$, to Bob
\begin{equation}
\label{AliceState}
\ket{\phi} = \alpha \ket{0}_{a_2} + \beta \ket{1}_{a_2}\ .
\end{equation}
The state of the whole system is the tensor product of the entangled qubits and Alice's private qubit
\begin{equation}
\label{teleStart}
\ket{\psi} = \ket{\phi} \otimes \ket{\theta} = \frac{1}{\sqrt{2}} (\alpha \ket{0}_{a_2} + \beta \ket{1}_{a_2})( \ket{0}_{a}\ket{0}_b + \ket{1}_{a}\ket{1}_b )\ . 
\end{equation}
Alices carries out the teleportation by first applying a CNOT gate to her two qubits, using $a_2$ as the control, followed by a Hadamard gate to her second qubit. It is straightforward to show that the resultant state of the system is
\begin{equation}
\label{teleportedState}
\begin{split}
\ket{\psi}' = \frac{1}{2} (\ &\ket{0}_{a_2}\ket{0}_{a} (\alpha \ket{0}_{b} + \beta \ket{1}_{b}) + \ket{0}_{a_2}\ket{1}_{a} (\alpha \ket{1}_{b} + \beta \ket{0}_{b})\ + \\ &\ket{1}_{a_2}\ket{0}_{a} (\alpha \ket{0}_{b} - \beta \ket{1}_{b}) + \ket{1}_{a_2}\ket{1}_{a} (\alpha \ket{1}_{b} - \beta \ket{0}_{b})\ )\ .
\end{split}
\end{equation}
Alice now measures the state of her qubits in the computational basis. It is clear that if she measures both qubits into state $\ket{0}$ then she has successfully teleported her qubit state to Bob. The other possible outcomes for Alice's measurement are straightforwardly dealt with.

This computation can be carried out with the TRS. We start by representing the starting state (\ref{teleStart}) as a term expression. To make the presentation more readable I will name parts of the term expression with $:=\ $. The shared Bell state is (\ref{BellState})
\begin{equation*}
\boldsymbol\theta := \mathtt{timesV}(\ \frac{1}{\sqrt{2}},\ \mathtt{plusV}(\ \mathtt{tensorV}(\ \vec{0}_a,\ \vec{0}_b\ ),\ \mathtt{tensorV}(\ \vec{1}_a,\ \vec{1}_b\ )\ )\ )\ ,
\end{equation*}
and Alice's state to teleport is (\ref{AliceState})
\begin{equation*}
\boldsymbol\phi := \mathtt{plusV}(\ \mathtt{timesV}(\ \alpha ,\ \vec{0}_{a_2}\ ),\ \mathtt{timesV}(\ \beta ,\ \vec{1}_{a_2}\ )\ )\ .
\end{equation*}
The starting state is the tensor product of these states
\begin{equation*}
\boldsymbol\psi := \mathtt{tensorV}(\ \boldsymbol\phi,\ \boldsymbol\theta\ )\ .
\end{equation*}

We need to define the action of the CNOT and Hadamard operators within the TRS. To do this we provide additional rules -- effectively teaching the TRS about the new operators. The Hadamard operator is defined by the rules
\begin{align}
\mathtt{apply}(\ \hat{h}_s,\ \vec{0}_s ) &\rightarrow \mathtt{timesV}(\ \frac{1}{\sqrt{2}},\ \mathtt{plusV}( \vec{0}_s, \vec{1}_s )\ )\ ,\\
\mathtt{apply}(\ \hat{h}_s,\ \vec{1}_s ) &\rightarrow \mathtt{timesV}(\ \frac{1}{\sqrt{2}},\ \mathtt{plusV}(\ \vec{0}_s,\ \mathtt{timesV}(-1, \vec{1}_s)\ )\ )\ ,
\end{align}
for any qubit space $s$. Notice that the operator is specified with respect to particular states (constant symbols, not variables) in a particular basis. In this problem we shall always work in the computational basis, but in more complex problems it is often desirable to work in several bases. In this case it would be necessary to adorn the vector symbols and rules with basis information, as well as providing rules for changing between bases.

The CNOT operator acts on a pair of qubits, in spaces $s_1$ and $s_2$, and has sort  $operator[s_1 \otimes s_2]$. It is defined by the rules
\begin{align}
\mathtt{apply}(\ \hat{c}_{s_1,s_2},\ \mathtt{tensorV}(\vec{0}_{s_1}, \vec{0}_{s_2})\ ) &\rightarrow \mathtt{tensorV}(\vec{0}_{s_1}, \vec{0}_{s_2})\ ,\\
\mathtt{apply}(\ \hat{c}_{s_1,s_2},\ \mathtt{tensorV}(\vec{0}_{s_1}, \vec{1}_{s_2})\ ) &\rightarrow \mathtt{tensorV}(\vec{0}_{s_1}, \vec{1}_{s_2})\ ,\\
\mathtt{apply}(\ \hat{c}_{s_1,s_2},\ \mathtt{tensorV}(\vec{1}_{s_1}, \vec{0}_{s_2})\ ) &\rightarrow \mathtt{tensorV}(\vec{1}_{s_1}, \vec{1}_{s_2})\ ,\\
\mathtt{apply}(\ \hat{c}_{s_1,s_2},\ \mathtt{tensorV}(\vec{1}_{s_1}, \vec{1}_{s_2})\ ) &\rightarrow \mathtt{tensorV}(\vec{1}_{s_1}, \vec{0}_{s_2})\ .
\end{align}

To construct the operators that act on the three-qubit system a third operator is needed -- the identity operator -- defined by the trivial rule
\begin{equation*}
\mathtt{apply}(\ \hat{id}_s,\ v_s\ ) \rightarrow v_s
\end{equation*}
for any space $s$ and any vector $v_s$.

Using these operators we can write the teleported state as
\begin{equation*}
\mathtt{apply}(\ \mathtt{compose}(\ \mathtt{tensorO}(\ \mathtt{tensorO}(\ \hat{h}_{a_2},\ \hat{id}_a,\ \hat{id}_b\ )\ ),\ \mathtt{tensorO}(\ \hat{c}_{a_2, a},\ \hat{id}_b\ )\ ),\ \boldsymbol\psi \ )\ .
\end{equation*}
Our task is to rewrite this expression using the rules defined above into the form of (\ref{teleportedState}). The details of this manipulation will not be presented as it involves 123 successive applications of the rules. In broad outline: the tensor product state is first expanded to a sum of tensor product basis states; the operator composition is then expanded and the CNOT operator applied; the Hadamard operator is applied and the resulting state is further expanded; finally, the terms are rearranged and collected to yield the result. The resulting term expression is
\begin{align*}
&\mathtt{timesV}(\ \frac{1}{2}\ ,\\
&\ \ \mathtt{plusV}(\ \mathtt{plusV}(\\
&\ \ \ \ \mathtt{tensorV}(\ \mathtt{tensorV}(\ \vec{1}_{a_2},\ \vec{0}_a\ ),
\ \mathtt{plusV}(\ \mathtt{timesV}(\ \alpha ,\ \vec{0}_b\ ),\ \mathtt{timesV}( - \beta,\ \vec{1}_b\ )\ )\ ),\\
&\ \ \mathtt{plusV}(\\
&\ \ \ \ \mathtt{tensorV}(\ \mathtt{tensorV}(\ \vec{0}_{a_2},\ \vec{1}_a\ ),
\ \mathtt{plusV}(\ \mathtt{timesV}(\ \alpha ,\ \vec{1}_b\ ),\ \mathtt{timesV}(\ \beta ,\ \vec{0}_b\ )\ )\ ),\\
&\ \ \ \ \mathtt{tensorV}(\ \mathtt{tensorV}(\ \vec{1}_{a_2},\ \vec{1}_a\ ),
\ \mathtt{plusV}(\ \mathtt{timesV}(\ \alpha ,\ \vec{1}_b\ ),\ \mathtt{timesV}(\ -\beta,\ \vec{0}_b\ )\ )\ )\ )\ ),\\
&\ \ \ \ \mathtt{tensorV}(\ \mathtt{tensorV}(\ \vec{0}_{a_2},\ \vec{0}_a\ ),
\ \mathtt{plusV}(\ \mathtt{timesV}(\ \alpha ,\ \vec{0}_b\ ),\ \mathtt{timesV}(\ \beta ,\ \vec{1}_b\ )\ )\ )\ )\ )\ .
\end{align*}
which can be seen to represent the Dirac-form expression (\ref{teleportedState}). It is perhaps surprising that a seemingly simple manipulation requires so many rule applications. Of the 123 applications 24 are of associativity and commutativity rules and could be eliminated by using a more sophisticated matching system. However, 99 rule applications still remain, the majority of which are of rules expanding tensor products of vector sums. It is testament to the power of mathematical notation in general, and Dirac's notation in particular, that so many formal operations can be performed so quickly by hand.

\section{Implementation}

It is desirable to implement the above described TRS in a mainstream computer algebra package. Calculations might then benefit from integration with the numerical, (scalar) analytic and visualization facilities of these packages. Most computer algebra packages offer some mechanism for implementing rewrite systems. However, as far as the author is aware, no general-purpose, mainstream package provides support for the algebraically parameterised sorts used in this paper. This section briefly describes a solution to this problem. Other details of the implementation are largely obvious\footnote{A proof-of-principle implementation may be obtained from the author's website\cite{jstar}.}.

A proof of principle implementation has been developed in the Mathematica \cite{Mathematica} computer algebra package. Mathematica has a sophisticated conditional rewriting engine, but offers no support for sorts -- all term expressions in Mathematica are of the same sort. It is straightforward, though, to add functions which reveal the sort of a term in the quantum algebra TRS. For convenience the problem is split into two parts. First, predicates are provided which indicate whether a term is a vector, operator, or scalar: \verb vectorQ  , \verb operatorQ  , and \verb scalarQ . The inductively defined term structure given in section \ref{terms} naturally suggests a recursive implementation. For example, \verb vectorQ  is defined as

\begin{samepage}
\begin{verbatim}
vectorQ = MatchQ[#,
     vector[__][__]
     | (plusV[ v1_?vectorQ, v2_?vectorQ ] 
	            /; ( spaceOf[v1] == spaceOf[v2] ))
     | timesV[ _?scalarQ, _?vectorQ ]
     | (apply[ o_?operatorQ, v_?vectorQ ] 
	            /; ( spaceOf[o] == spaceOf[v] ))
     | tensorV[ _?vectorQ, _?vectorQ ]
    ] &;
\end{verbatim}
\end{samepage}
Note that vector and operator constants are indicated by expressions that have head \verb vector  and \verb operator  respectively. The first curried argument of a vector or operator constant is used to indicate the space to which it belongs. The second argument (or argument list) names the constant, and can carry ancillary information, for instance about the basis to which the name refers.

Second, a function is provided that returns the Hilbert space to which a vector or operator belongs. For brevity, only a few clauses of the function are presented to indicate the general idea

\begin{samepage}
\begin{verbatim}
spaceOf[vector[a_][__]] := a
spaceOf[timesV[s_?scalarQ,v_?vectorQ]]:= spaceOf[v]
spaceOf[apply[o_?operatorQ,v_?vectorQ]] := sameSpace[o,v]
spaceOf[tensorV[v1_?vectorQ,v2_?vectorQ]] := 
                     tensorSpace[spaceOf[v1],spaceOf[v2]]
\end{verbatim}
\end{samepage}
The function \verb sameSpace  returns the space its two arguments belong to, if they are the same, otherwise generating an error. Of note is the function \verb tensorSpace . It is this function that ensures that the sorts respect the semi-group structure of the Hilbert space tensor product. Its implementation is simple
\begin{verbatim}
tensorSpace[s1_, s2_] := Sort[Join[s1,s2]]
\end{verbatim}
with the set union providing the required algebraic structure. Rules of the TRS are implemented as conditional rewrite rules, conditioned on appropriate results of these functions that reveal the sort of a term.

This implementation is extremely inefficient and makes no attempt to integrate with the general-purpose algebraic functions of Mathematica. Furthermore, error checking is incomplete as a term's sort is only checked when a conditional rewrite rule is applied. It is an open research question how to best develop a well-integrated, efficient implementation of the system in a general-purpose computer algebra package.

\section{Conclusion}

I have described the mathematical framework for building a computer algebra system for representation-invariant quantum mechanical calculations. This framework significantly differs from all previous work by capturing the underlying structure of the mathematics of quantum mechanics, rather than directly capturing the most commonly used notation. I have demonstrated by example that the framework can be used to solve non-trivial problems.

A number of research directions are suggested by this work. Further formal characterisation of the TRS would be of interest; in particular determining whether the TRS has the desirable properties of termination and confluence. Integration of the system into a general-purpose computer algebra system also requires further work. The unsorted rewrite-systems that most packages are built on present a particular problem for the implementation of complex algebraic systems of the type described herein.

I would like to thank Mike Tarbutt and Ben Sauer for their comments on this manuscript. The author is supported by an STFC Advanced Fellowship.

\end{document}